# Efficiency limits of Perovskite Solar Cells with Transition Metal Oxides as Hole Transport Layers

Dhyana Sivadas, Swasti Bhatia and Pradeep R. Nair

Dept. of Electrical Engineering, Indian Institute of Technology Bombay, Mumbai, India

Transition metal oxides (TMOs) like MoOx are increasingly explored as hole transport layers for perovskite based solar cells. Due to their large work function, the hole collection mechanism of such solar cells are fundamentally different from other materials like PEDOT: PSS and the associated device optimizations are not well elucidated. In addition, the prospects of such architectures against the challenges posed by ion migration are yet to be explored – which we critically examine in this contribution through detailed numerical simulations. Curiously, we find that, for similar ion densities and interface recombination velocities, ion migration is more detrimental for Perovskite solar cells with TMO contact layers with much lower achieveable efficiency limits (~21%). The insights shared by this work should be of broad interest to the community in terms of long term stability, efficiency degradation and hence could help critically evaluate the promises and prospects of TMOs as hole contact layers for perovskite solar cells.

## I. INTRODUCTION

Recent perovskite solar cells report efficiencies comparable to the market leader c-Si solar cells[1]. Given this excellent improvement, current emphasis is more on stability and possible pathways to commercialization[2,3]. A combination of material/interface engineering is one among the many prominent options being explored[4–6]. To mitigate the shortcomings associated with organic materials, Transition metal oxides (TMO) like MoOx, WOx are considered as possible candidates for hole transport layers[7,8] (see Fig. 1a). Indeed, high transparency, ease of fabrication, high work function, non- toxicity and stability in ambient conditions are some of the favorable characteristics of TMO[9]. Still, given the fundamental difference in their hole collection/quenching properties[10], it is not evident whether such solar cells could be as good as the traditional architectures. Further, the impact of ion migration on this new class of solar cells is yet to be elucidated. Given this, here we focus on the efficiency limits and influence of ion migration on perovskite solar cells with TMO as hole transport layers.

TMOs as hole transport layer and/or inter layer in perovskite solar cells has been explored by multiple groups. For example, Tseng et al., used TMO as a hole transport layer to replace PEDOT:PSS in perovskite based solar cells[8]. With F4-TCNQ modification in MoOx, Chen et al., reported enhanced efficiency of 16.26% for perovskite solar cell[11]. Besides, MoOx is also used as interlayer between perovskite and ETL to improve the performance of perovskite solar cell[7,12,13]. Not surprisingly, TMOs are already widely explored in organic devices. For example, Tokito et al.,[14] used TMO as an interlayer in organic light emitting diode, which was then explored in the organic photovoltaic devices[15,16]. Further, Si based carrier selective solar cells also use TMOs as hole transport layers [10,17–20].

A quick review of the above literature reveals many relevant unanswered questions. For example, (i) What would be efficiency limits of perovskite solar cells with TMO as hole collection/transport layer? (ii) Is the new architecture resilient to ion migration and interface degradation, (iii) What are the device optimization pathways to improve the performance? In this context, through detailed modeling and simulations, here we show that Perovskite solar cells with TMO as hole transport/extraction layers are significantly affected by bias dependent carrier collection phenomena (along with ion migration and interface recombination) which limits achievable efficiencies to the order of 21% while devices with conventional architectures and similar parameters could yield above 25% efficiency. Below, we describe the model system, simulation methodology, results and their implications.

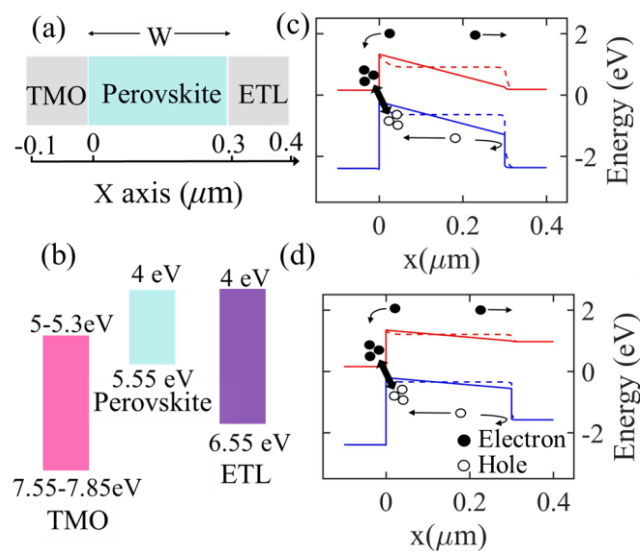

**FIG. 1.** Model system. (a)Schematic of the device used for simulations, (b) Energy levels of the materials, (c) Energy band (EB) diagram of TPSC near short circuit condition, (d) near open circuit condition. (Dashed line depicts the equilibrium EB diagram in the presence of ions while solid lines denote EB diagram in the absence of ions.).

## II. MODEL SYSTEM

Figure 1a shows a schematic of the device under consideration. Here, intrinsic perovskite is sandwiched between Transition metal oxide (TMO) and Electron transport layer (ETL). The TMO acts as the hole transport layer. Interestingly, both the TMO and the ETL are n-type doped – in contrast to the usual p-type doping used for HTL in conventional perovskite solar cells. The work function[8,21,22] of TMO of range 5.0-5.5eV whereas ETL band alignment is such that it perfectly blocks holes and collects electrons (see Fig.1b). Note that this is a NIN structure in contrast to the conventional perovskite solar cells (CPSC) which are essentially PIN structures[23–25](see Fig. S1, Suppl. Mat.). As per this terminology, n-doped TMOs with large work function like MoOx, WOx could function as TPSC, while NiOx could act as a traditional hole transport layer (due to its unique band alignment properties)[26,27].

An important performance limiting parameter of perovskite solar cells is the presence of ions in the active layer[28,29]. To fully explore the implications of ion migration in a NIN architecture, we developed a self-consistent drift diffusion model, as described below. The model accounts for electrostatics and continuity equations with explicit consideration of ions (negative mobile ions and positively charged immobile ions - in accordance with literature[30,31]). Specifically, the equations for the perovskite region are:

$$\frac{d^2V}{dx^2} = -\frac{q}{\varepsilon}\left(p - n + N_{dop} + N_{I,p} - N_{I,n}\right) \qquad (1)$$

$$\frac{1}{q}\frac{\partial J_n(x)}{\partial x} + G - R = 0 \qquad (2)$$

$$-\frac{1}{q}\frac{\partial J_p(x)}{\partial x} + G - R = 0 \qquad (3)$$

Here $V$ is the potential, q is the charge, $\varepsilon$ is the dielectric permittivity, $n$ is the electron density, $p$ is the hole density, $x$ is the spatial coordinate, $J_n$ is the electron current and $J_p$ is the hole current. $N_{dop}$ denotes doping concentration of the material and $N_I$ denotes the ion density (assumed to be present only in the perovskite active region. Subscript p or n denotes positive or negative ions, respectively). G is the generation rate inside perovskite and is taken as $5.21 \times 10^{21}$ $cm^{-3}/s$ under 1 Sun illumination (which results in a $J_{sc} \sim 25 mA/cm^2$). Recombination (R) is modelled by SRH (parameter, $\tau$), Auger (parameter $C_A$) and Radiative recombination (parameter, $C_R$).

We assume that the ions are confined to the perovskite active layer, and that the active layer is charge neutral with respect to ions. Such a charge neutrality principle is succinctly represented as

$$\int_0^W \left(N_{I,p}(x) - N_{I,n}(x)\right) dx = 0 \qquad (4)$$

where W is the thickness of the perovskite active layer (see Fig. 1a). Note that for positive immobile ions, we have $N_{I,p}(x) = I_0$, where $I_0$ is a constant which denote the average ionic density. In the absence of any generation/annihilation of ions and using Einstein relation, analytical solutions to steady state continuity equations for ions indicate that negative mobile ions vary exponentially with electrostatic potential (i.e., $N_{I,n}(x) = K e^{\frac{V(x)}{V_T}}$, where K is prefactor and $V_T$ is the thermal voltage. This is nothing but the Boltzmann distribution itself). Accordingly, the governing equation for mobile ion distribution is

$$I_0 \times W - \int_0^W K e^{V(x)/V_T} dx = 0 \qquad (5)$$

Note that the prefactor K varies with applied bias. Equation (5) obviates the need to write a separate continuity equation for mobile ions (as it directly uses the solution of ionic continuity equation). Accordingly, the steady state performance of perovskite solar cells could be obtained through self-consistent solution of equations (1)-(3) and (5).

In this model, we explicitly consider carrier recombination at material interfaces in addition to the various bulk recombination mechanisms (trap assisted SRH, radiative and Auger; parameters provided in Table S1, supplementary material). At ETL/Perovskite and HTL/Perovskite interfaces, carrier loss is accounted through interface recombination velocities ($S_v$). In addition, band-band carrier recombination across interfaces ($R_{INT}$) is considered at TMO/Perovskite interface (see the band diagrams in Figs. 1c and 1d) which is modeled as,

$$R_{INT}(x=0) = B(n_s p_s - n_{s,0} p_{s,0}) \qquad (6)$$

where B is the hole extraction parameter (assumed as $10^{-9}$ $cm^3/s$, in accordance with literature[32]). Note that $n_s$ and $p_s$ denote the respective carrier densities at the interface. Specifically, $n_s = n(x = 0^-)$, the electron density in TMO at the interface (see the schematic, Fig. 1a) while $p_s = p(x = 0^+)$, the hole density in perovskite at the interface. The subscript 0 for terms in eq. (6) denotes the equilibrium values. The hole extraction mechanism described by eq. (6) is central to the operation of perovskite solar cells with large work function TMOs as hole transport layers. This mechanism allows recombination between photogenerated holes in perovskite and electrons in TMO – which enables extraction/collection of photo-generated holes by the TMO.

The above equations are self consistently solved using finite difference method with Scharfetter-Gummel discretization and Newton's method for iterative solution of coupled non-linear equations[33,34]. Other parameters used in simulations are provided in Table S1 of the supplementary material. Note that our simulations are calibrated against the experimental data (see Fig. S2, Suppl. Mat.).

## III. RESULTS AND DISCUSSION

The energy band (EB) diagrams for TPSC at short circuit condition and near open circuit conditions are provided in parts (c) and (d) of Fig. 1. Corresponding band diagrams for CPSC are provided in the supplementary material (see Fig. S1). A comparative study of the same reveals several interesting aspects. Specifically, we find:

(i) The electron blocking at the TMO/Perovskite interface is due to the band bending in Perovskite (see Fig. 1c). At near open circuit conditions, the same electron blocking barrier is almost non-existent (see Fig. 1d). In comparison, electron blocking at HTL is more effective in CPSC (due to the offset in conduction band, see Fig. S1, Suppl. Mat.).

(ii) Negligible hole injection occurs in TPSC under dark conditions (as it is a NIN structure),

(iii) For TPSC, applied forward biases reduce the barrier for electron injection (from ETL to TMO) and hence over-the-barrier electron current from ETL to TMO is expected to be the dominant dark current component (see Fig. 1d). The same could significantly limit the $V_{oc}$ of TPSCs.

(iv) Photo-generated holes in TPSC are trapped in a potential well due to the barriers in the Valence band. Hence, the recombination mechanism described by eq. (6) is crucial in collection of such photo-generated holes by the TMO.

The above inferences could be significantly influenced by the presence of mobile ions in perovskite region. It is well known that mobile ions screen the built-in field (due to $V_{BI}$) of the device and reduce the internal electric field[35,36] thus affecting efficiency in CPSC (Fig. S1, see the supplementary material). However, the influence of ions on the performance of TPSC is yet to be clearly elucidated. The band diagram shown in Figs. 1c and 1d indicate that similar concerns are valid for TPSCs as well. In particular, Fig. 1c indicates that the potential barrier which reduces the electron flow to TMO is greatly reduced in the presence of high density of mobile ions. Accordingly, significant bias dependent photo carrier collection and hence reduced fill factor (FF) is expected due to ion migration effects.

The distinct band level alignments and different hole extraction mechanism in TPSCs result in starkly different characteristics during illumination. For example, Fig. 2 shows the current–voltage (J-V) characteristics of TPSC in the presence and absence of ions. A much-reduced current extraction is seen at high biases for TPSC with ions. Also, in the presence of ions, a deviation from the superposition[37] between dark and light JV is observed for TPSC (as compared to CPSC, see Fig. S3, Suppl. Mat.). This implies that in the presence of ions, photocurrent is more bias dependent in TPSC unlike CPSC.

To explore further, the spatial profiles and bias dependence of electron ($J_n$) and hole current ($J_p$) components of TPSC are plotted in Fig. 3. Specifically, we find that the entire photo-generated carriers at $SC$ conditions are collected at respective contacts ($|J_{tot}| = |J_{sc}| = 25\,mA/cm^2$, see Fig. 3a). In addition, both the electron and hole current components are in the −'ve x direction over the entire active layer thickness (see red and blue curves in Fig. 3a). Further, there is no hole current component in the TMO (the blue curve goes to zero for $x < 0\mu m$). Indeed, the entire electron current component in TMO perfectly matches the hole current component from the perovskite at the TMO/Perovskite interface at short circuit conditions (i.e., $J_p(x = 0^+) = J_n(x = 0^-)$) – which is nothing but extraction of photo-generated holes in perovskite by the electrons in TMO (made possible through the recombination mechanism described by eq. (6)).

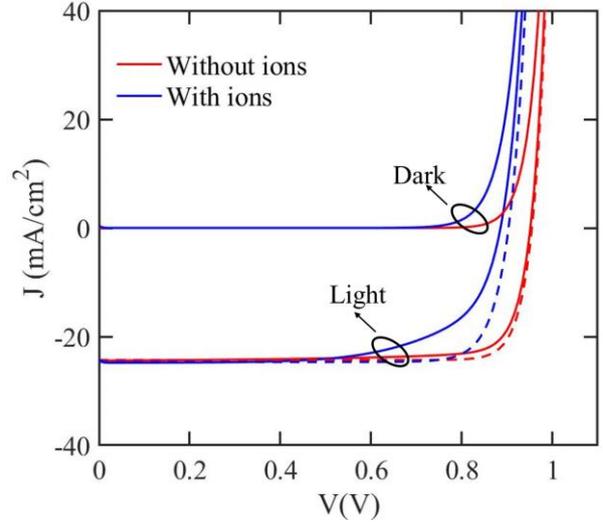

**FIG. 2.** Current-voltage (J-V) characteristics of TPSC in the absence and presence of ions (ion density=$5 \times 10^{17} cm^{-3}$). Dashed line indicates downshifted dark JV characteristics (i.e., $J_{dark} - J_{SC}$). Both devices have $V_{BI}$ of 1.2V.

Figure 3b shows the bias dependence of these individual current components. Interestingly, $J_p(x = 0^+)$ is almost invariant with the bias (see blue curve in Fig. 3b) – which indicates that photo-generated holes are indeed extracted at the TMO/perovskite interface. So, the bias dependence of photo-current is not contributed by the extraction of holes at TMO/perovskite interface. Now the only unknown is the electronic component which results in extraction of photo-generated holes – whether these electrons are from the TMO or not? Curiously, Fig. 3b shows that $J_n(x = 0^+)$ increases monotonically with applied bias. Note that a positive value of current indicates a current flowing in the +'ve x direction. Hence the electrons contributing to $J_n(x = 0^+)$ actually flow towards the TMO/perovskite interface (i.e., electrons flow in the −'ve x direction which results in an electronic current in the +'ve x direction). We know that the external current ($J_{Tot}$) could be only due to the electrons that flow from the bulk of TMO and Fig. 3b shows that $J_{Tot} = J_n(x = 0^-)$ for the entire range of applied bias. However, $J_n(x = 0^-) \neq J_p(x = 0^+)$ for large applied biases. In addition, $J_n(x = 0^-) = J_n(x = 0^+) + J_p(x = 0^+)$. All these indicate that, at large applied biases, a significant portion of the photo-generated electrons from perovskite cross over to the TMO and recombines with the photo-generated holes at TMO/perovskite interface – i.e., the very mechanism that allows hole extraction at $SC$ conditions contribute to bias dependent photo collection and poor FF, thus limiting achievable efficiency. The bias dependent reduction of barrier which prevents the flow of electrons to TMO with applied bias is shown in Fig. S4, Suppl. Mat.

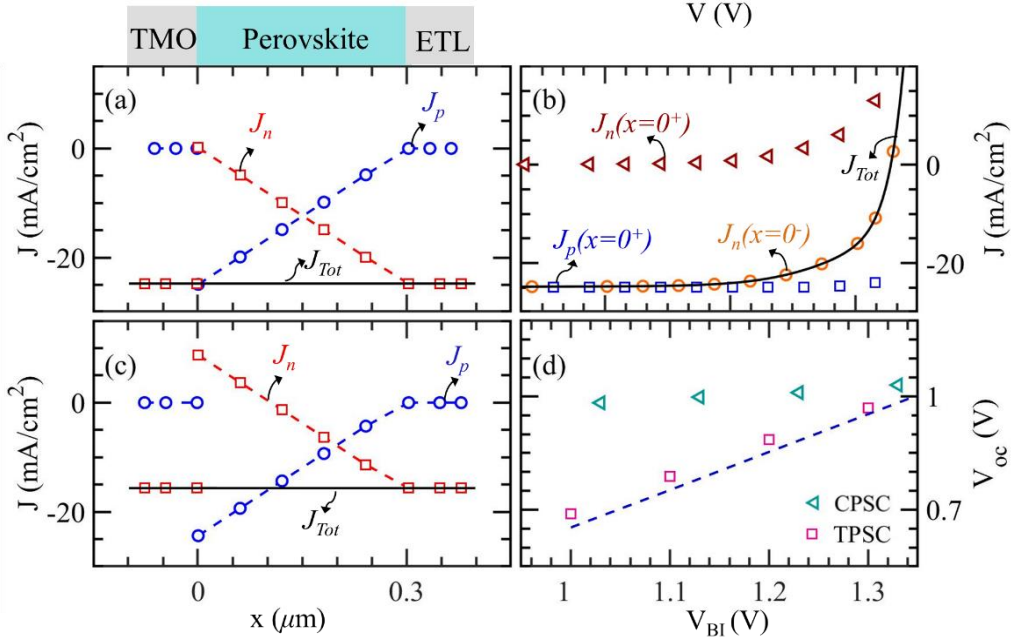

**FIG. 3.** Spatial and bias dependence of photo-current components. (a) Spatial profiles of electron ($J_n$) and hole current ($J_P$) components of TPSC at $SC$, (b) Bias dependence of current components at perovskite-TMO interface, (c) Spatial profiles of electron ($J_n$) and hole current ($J_P$) components of TPSC near $OC$ ($V_{bias} = V_{OC} - 85mV$), (d) Variation of open circuit voltage with built-in potential for CPSC and TPSC. The dashed lines indicate theoretical estimates (using eq. (7)). These simulations are for ion density of $5 \times 10^{17} cm^{-3}$.

The above inferences are supported by the results at near $OC$ conditions shown in Fig. 3c. Here, we find that at higher biases the electrons move in both directions (see Fig. 3c). This is in contrary to $SC$ conditions where the photogenerated electrons move only towards ETL (as shown in Fig. 3a). Still, for both SC and near $OC$ scenarios, we observe that the total current is spatially constant which reflects well on the accuracy of simulations. Here, we find that while the hole current predominantly flows in the –'ve x direction (see blue curve in Fig. 3c), the electron current, curiously and in contrast to that of the $SC$ conditions, flows in the +'ve x direction over a significant region of the active material (see red curve in Fig. 3c). Hence, as discussed in previous paragraph, photo-generated electrons cross over to TMO and recombine at the interface resulting in efficiency loss. A close analysis of the recombination profiles at near $OC$ condition (see Fig. S5, Suppl. Mat.) also indicates that the dominant component of the carrier loss is through the cross-recombination facilitated by eq (6).

Figure 3d shows the variation of open circuit voltage ($V_{OC}$) with the built-in potential ($V_{BI}$) for CPSC and TPSC devices. Here, while the $V_{OC}$ of CPSC is larger and relatively independent of $V_{BI}$, the $V_{oc}$ of TMO-solar cell, curiously, varies linearly with $V_{BI}$. The loss of photogenerated electrons to the TMO layer due to the absence of conduction band offset is the reason for the inferior $V_{oc}$ of TPSC. An estimate for the same could be obtained as follows:

We know that $J_{Tot} = J_n(x = 0^+) + J_P(x = 0^+)$. Hence, by definition, $V_{OC}$ is nothing but the voltage when $J_n(x = 0^+)$ equals $J_P(x = 0^+)$. Note that both the dark injection as well as the bias dependent photo-current contributes to $J_n(x = 0^+)$. Analytical estimates for the dark injection current are not available for devices in the presence of ion migration. However, dark injection current for NIN diodes in the absence of any ions can be easily derived (Section VI, Suppl. Mat.) and is given as

$$J_{dark} = -\frac{qN_{TMO}e^{\frac{-V_{BI}}{kT}}\mu_n\left(\frac{V_{BI}-V}{W}\right)}{\left(e^{\frac{V-V_{BI}}{V_T}} - 1\right)}\left(e^{\frac{V}{V_T}} - 1\right) \qquad (7)$$

where $N_{TMO}$ is the doping concentration of TMO layer. A first order estimate for $V_{OC}$ can be obtained by equating $J_{dark}$ to $J_{SC}$. Interestingly, this model accurately predicts the $V_{BI}$ dependence of $V_{OC}$ in TPSC devices (see blue dashed lines in Fig. 3d). Section V of supplementary material provides analysis related to detailed balance $V_{OC}(\sim 1V)$ for CPSC devices.

Having clearly identified the performance limiting mechanism in TPSC, let us now estimate achievable efficiency limits. The bias dependence of photo-current is also influenced by parameters like ion migration and interface recombination ($S_v$). Given the quality of perovskite material (carrier lifetime of the order of $1\mu s$), the carrier collection lengths are larger than active layer thickness – irrespective of whether drift or diffusion-based carrier collection scheme. Hence, we expect $J_{sc}$ to be rather invariant with ion density and $S_v$ (unless limited by very large $S_v$). Further, the $V_{oc}$ of these cells could be limited by bias dependent photo-carrier collection which is critically influenced by $V_{BI}$ (see Fig. 3d) and, hence little dependence on ion density. The FF, which indicates the photo-carrier collection efficiency at maximum power conditions, is influenced by the potential barrier that prevents carrier collection at wrong electrodes. As indicated by the E-B diagrams (see Fig. 1), the potential barrier which allows such selective collection of carriers decreases with ion density and hence the FF is expected to significantly decrease with ion density. Taking these insights into account, we expect the efficiency of TPSCs to be lower than CPSC and to be significantly influenced by ion migration and interface recombination.

Figure 4 shows a comparison of TPSC and CPSC in the presence of ion migration and interface recombination (at perovskite/transport layer interface). Fig. 4a indicates that the performance of CPSC degrades as a function ion density – in accordance with literature[38,39]. For same set of parameters, we find that TPSC efficiency is lower than that of CPSC. Further, with an increase in ion density, efficiency degradation in TPSC is more as compared to CPSC. The E-B diagrams in Fig. 1 indicate that the fraction of $V_{BI}$ that drops in the Perovskite reduces in the presence of large ionic density, (i.e., the net band bending in Perovskite active layer. The rest of $V_{BI}$ will be dropped across the transport layers, see Fig. 1c). Accordingly, electron blocking barrier at the TMO/Perovskite interface reduces significantly with ion density leading to the trends shown in Fig. 4a (see Fig. S6, Suppl. Mat. for the influence of ion density on EB diagrams). Naturally, an increase in $V_{BI}$ is expected to improve TPSC performance – as supported by the results in Fig. 4a.

The abovementioned decrease in perovskite band bending has significant implications towards carrier loss due to interface recombination. As the perovskite band bending reduces, carrier recombination through non-radiative recombination at the perovskite/transport layer interface increases. In addition to the carrier loss due to trap assisted recombination, the reduction of electron blocking restricts the efficiency of TPSC to be much lower than that of CPSC. This is seen in Fig. 4b where the initial efficiency values are much lower for TPSC. We also find that the efficiency is rather independent of $S_v$ (till $10^3 cm/s$) in TPSC. Due to the high work function of TMO, one type of carrier is depleted inside perovskite near the interface and accordingly the recombination is not significant at TMO/Perovskite interface (see Fig. S7, Suppl. Mat.). Hence, although the absolute values are lower, the efficiency loss with $S_v$ is not significant for TPSC (for rather small $S_v$). For large $S_v$, both TPSC and CPSC show similar efficiencies. The variation of individual performance parameters related to Fig. 4 is provided in section X of supplementary material. In addition, our simulations indicate that, for heavily doped TMOs as HTL, TPSC efficiencies are unaffected for a broad range of B (hole extraction parameter, see Fig. S10, Suppl. Mat.). This is not surprising as the for the parameters involved, the recombination in eq. (6) is indeed limited by the hole availability.

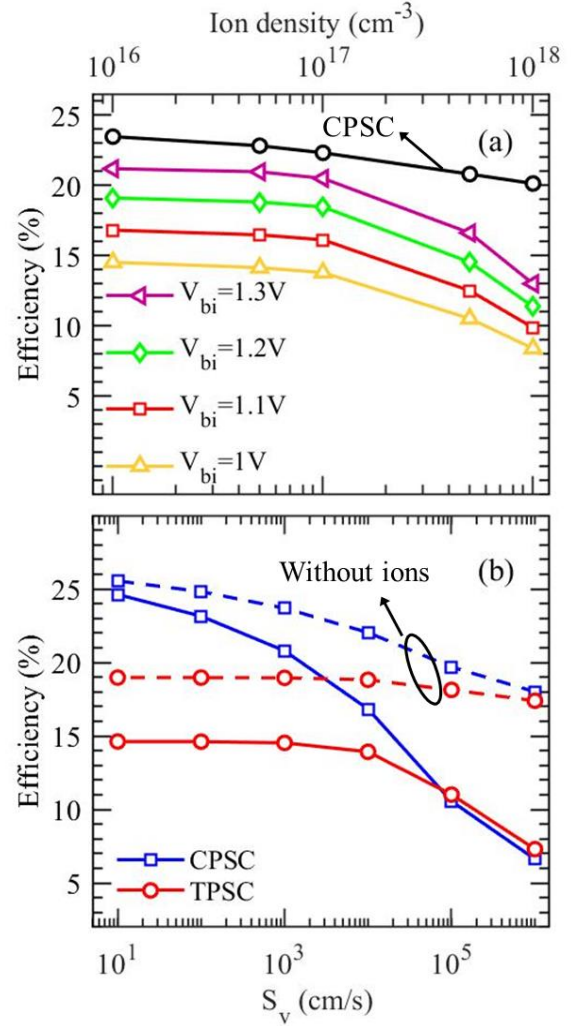

**FIG. 4.** Influence of ion migration and interface recombination on the efficiency of perovskite solar cells. (a) Effect of ion density on the performance of TPSC for different $V_{bi}$ (for comparison, CPSC results are for $V_{BI} = 1.2V$), (b) Effect of surface recombination velocity on the performance of TPSC and CPSC in the presence of ions (ion density=$5 \times 10^{17} cm^{-3}$, $V_{bi} = 1.2V$). Dashed line indicates results in the absence of ions.

Finally, it is well known that CPSC devices report efficiencies of the order of 25% with improved current collection schemes[1]. In this regard, what might be the best performance that one could achieve using TPSC? We note that while the best efficiency corresponding to $V_{BI} = 1.2V$ and $S_v = 10 cm/s$ is ~25% for CPSC, the corresponding efficiency is only ~19% for TPSC (see Fig. 4b). However, an increase in the $V_{BI}$ improves the open circuit voltage and thereby increase efficiency of TPSC considerably. Hence, devices with $V_{BI} = 1.3V$ and $S_v = 10 cm/s$ gives an efficiency of ~21% (see Fig. S11, Suppl. Mat.). Additional electron blocking phenomena – through fixed charge or band offsets could improve the performance of TPSCs significantly. In the absence of the same, practical achievable efficiency of TPSC will be of the order of 21%. Note that this conclusion is applicable only for devices with TMOs like MoO$_x$, WO$_x$, etc. (i.e., with NIN architecture) while NiO$_x$ with its unique band level alignment could yield performance similar to CPSCs. Additional aspects from long-term stability could be appealing for this architecture.

## IV. CONCLUSION

In summary, we identified the fundamental charge transport mechanism and the origin of bias dependent photocurrent in TMO based perovskite solar cells (TPSC). Our results reveal that, insufficient electron blocking in the TMO limits the performance of TPSC. As a result, the ion migration inside the perovskite causes a drastic reduction of efficiency in TPSC compared to the conventional pin solar cell (CPSC). Conversely, although the efficiencies are lower, TPSC suffers from lower relative performance loss due to the interface degradation. Moreover, as the inferior performance of TPSC is primarily due to the over-the barrier electron current, interface engineering like additional electron blocking layer or fixed charge could indeed help improve the performance significantly. We hope that further research efforts on this new architecture could overcome the shortcomings and pave the way to the eventual commercialization.

## SUPPLEMENTARY MATERIAL

See the supplementary material for CPSC model system, detailed derivation of dark current and open circuit voltage, explanation regarding effect of different parameters (ions, surface degradation, B parameter) on TPSC.

## ACKNOWLEDGMENTS

This project is funded by Science and Engineering Research Board (SERB, project code: CRG/2019/003163), Department of Science and Technology (DST), India. Authors acknowledge IITBNF and NCPRE for computational facilities. PRN acknowledges Visvesvaraya Young Faculty Fellowship.

## DATA AVAILABILITY

The data that supports the findings of this study are available within the article [and its supplementary material].

# Efficiency limits of Perovskite Solar Cells with Transition Metal Oxides as Hole Transport Layers

*Dhyana S, Swasti Bhatia and Pradeep R. Nair*

Department of Electrical Engineering, Indian Institute of Technology Bombay, Powai, Mumbai, 400076,

India

**E**mail: dhyana@iitb.ac.in, email: prnair@ee.iitb.ac.in

This document contains the following additional material to support the discussions in the main manuscript:

I.      Model System.

II.     Calibration against Experimental Result.

III.    Superposition of light and dark J-V characteristics in CPSC.

IV.    Energy band diagram of TPSC at different biases.

V.     Recombination profiles of TPSC near open circuit condition.

VI.    Derivation of dark current in TPSC.

VII.   Estimation of open circuit voltage.

VIII.  Energy band diagram of TPSC in the presence of ions.

IX.    SRH profile near open circuit condition.

X.     Effect of Ion density and Surface recombination velocity in TPSC.

XI.    Effect of B parameter on JV characteristics of TPSC.

XII.   JV characteristics of TPSC with efficiency of 21%.

**Section I: Model System**

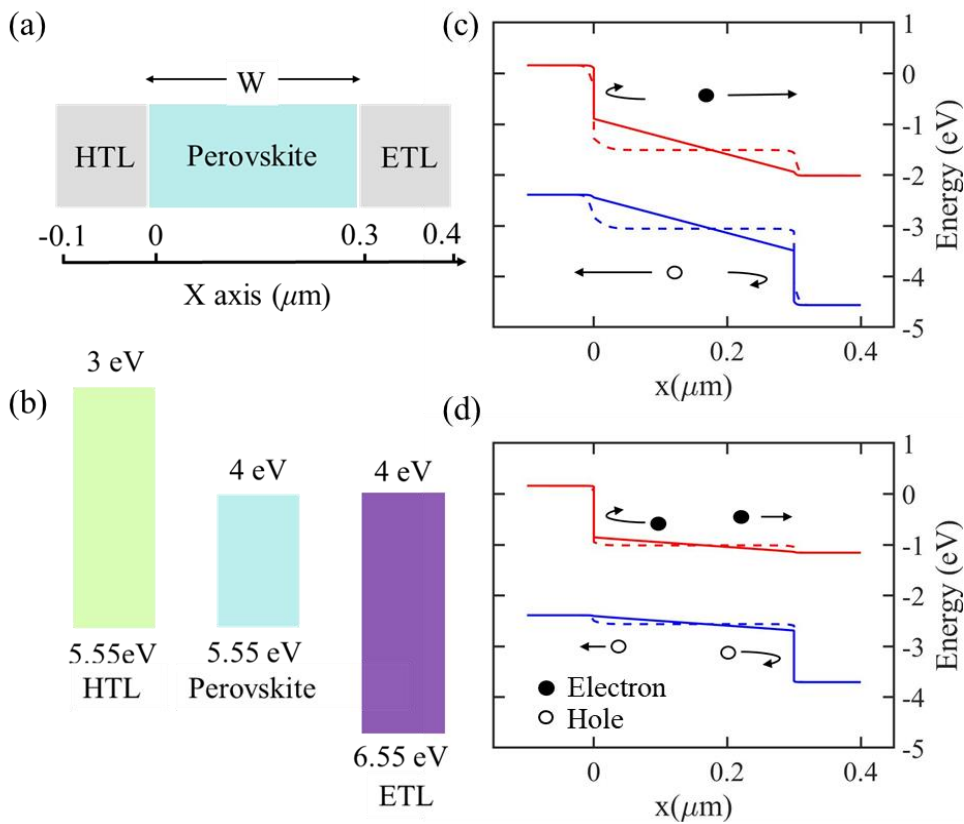

**FIG. S1.** (a)Schematic diagram of the CPSC device used for simulation, (b) Energy levels of the materials, (c) Energy band diagram of CPSC near short circuit condition, (d)near open circuit condition. Here, dashed line depicts the energy band diagram in the presence of ions. As mentioned in the main manuscript the electric field inside the perovskite is reduced significantly in the presence of ions. Further, the schematic of in the main manuscript is shown in Table S1[1]. The parameters used for the simulation as discussed

| Parameter | TMO | Perovskite | ETL |
|---|---|---|---|
| Thickness (nm) | 100 | 300 | 100 |
| Electron Affinity, EA (eV) | 5-5.32 | 4 | 4 |
| Band gap, Eg (eV) | 2.55 | 1.55 | 2.55 |
| Effective density of states, Nc, Nv ($cm^{-3}$) | $5 \times 10^{20}$, $5 \times 10^{20}$ | $10^{19}, 10^{19}$ | $5 \times 10^{20}$, $5 \times 10^{20}$ |
| Doping, $N_{dop}$ ($cm^{-3}$) | $10^{18}$ | - | $10^{18}$ |
| Mobility (n/p), $\mu$ ($cm^2/Vs$) | $12 \times 10^{-3}$ | 10 | $12 \times 10^{-3}$ |
| Recombination Parameters | $\tau = 10^{-6}$ | $\tau = 10^{-6}$<br>$C_R = 3 \times 10^{-11}$<br>$C_A = 10^{-29}$ | $\tau = 10^{-6}$ |
| Hole extraction parameter B ($cm^3/s$) | - | $10^{-9}$ | - |

**Table S1.** Parameters used for numerical simulation.

## Section II: Calibration against Experimental Result[2]

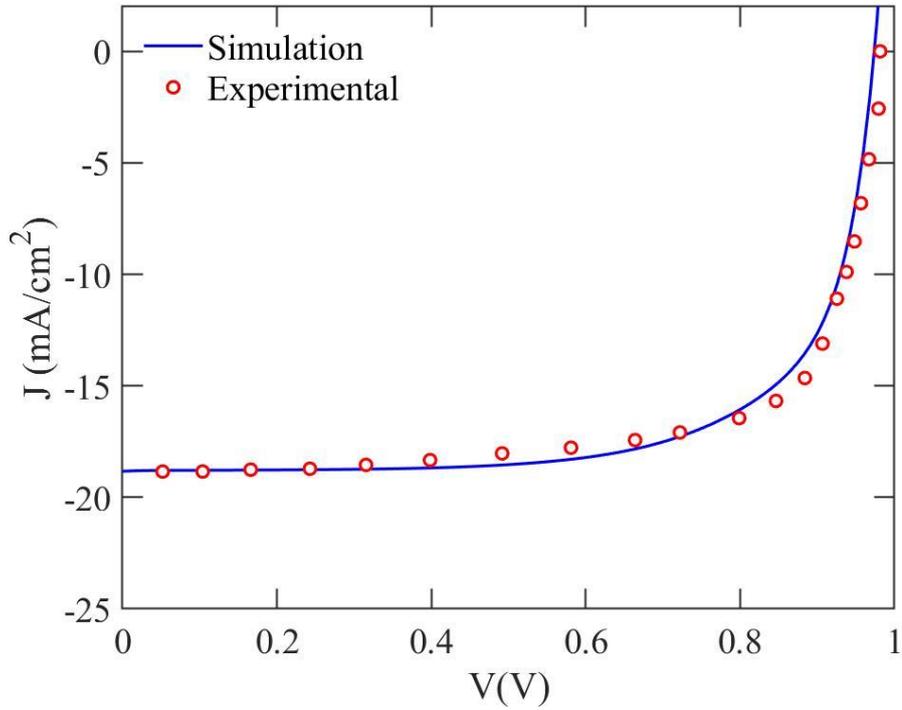

**FIG. S2.** J-V Characteristics of TPSC obtained from simulation is compared with the experimental data. Here, we considered the work function of TMO as 5.48eV, SRH life time ($\tau$) of active layer as $7.2 \times 10^{-8} s$ and ion density inside the active layer as $5 \times 10^{17} cm^{-3}$.

**Section III: Superposition of light and dark J-V characteristics in CPSC**

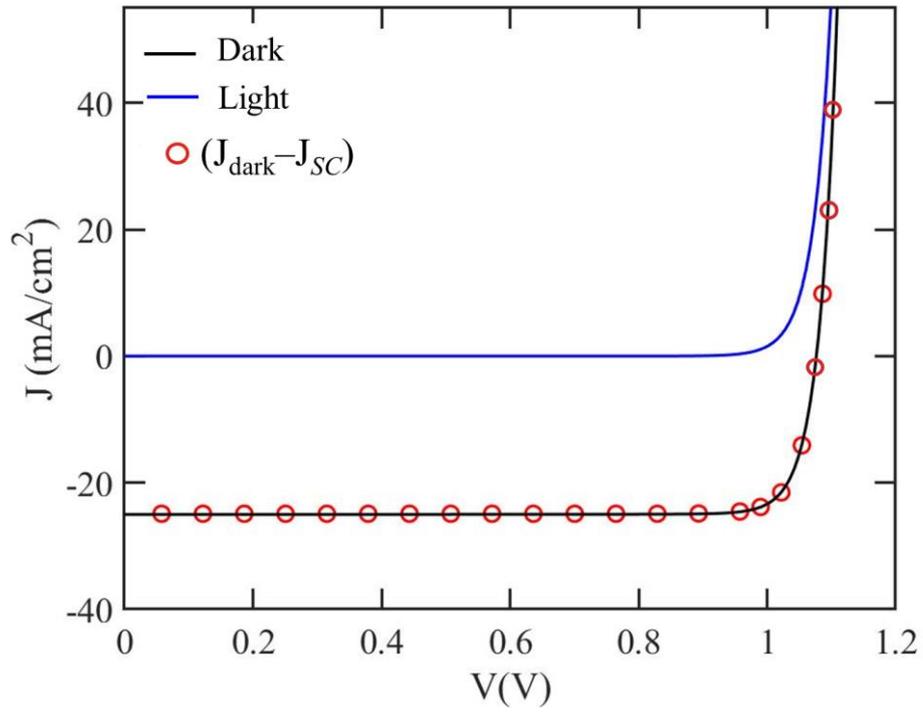

**FIG. S3.** Light and dark current-voltage (J-V) characteristics of CPSC in the presence of ions (ion density=$5 \times 10^{17} cm^{-3}$). Data set with red symbols indicate superposition of dark and photo-current ($J_{dark} - J_{SC}$).

**Section IV: Energy band diagram of at different biases**

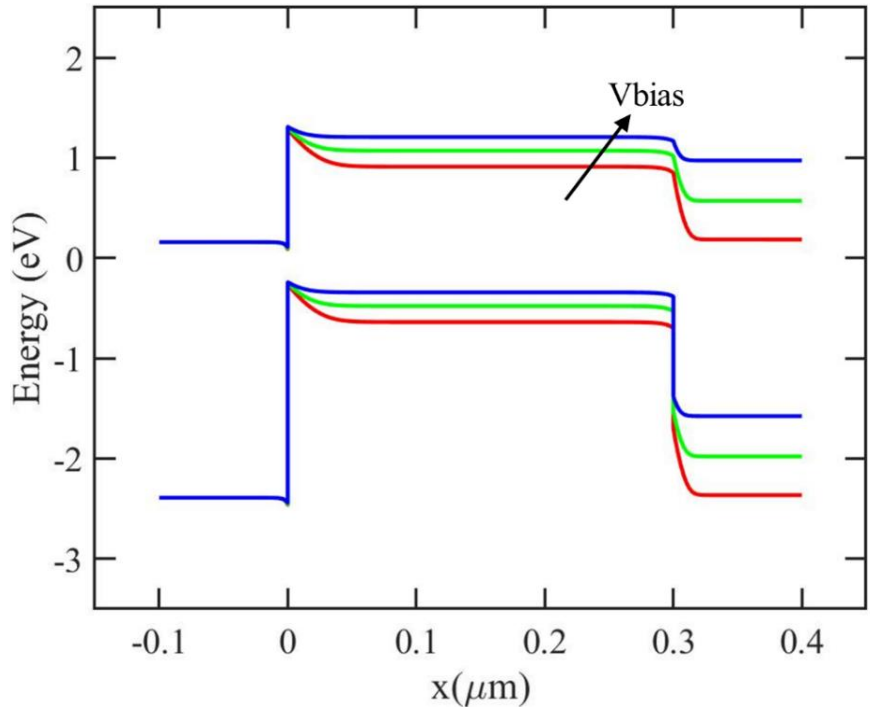

**FIG. S4.** Energy band diagram of TPSC in the presnce of ions (ion density=$5 \times 10^{17} cm^{-3}$) at different biases. Indeed the barrier for electron reduces with applied bias which helps the photogenerated electrons to cross over to the TMO as discussed in the main manuscript.

**Section V: Recombination profile of TPSC near open circuit condition**

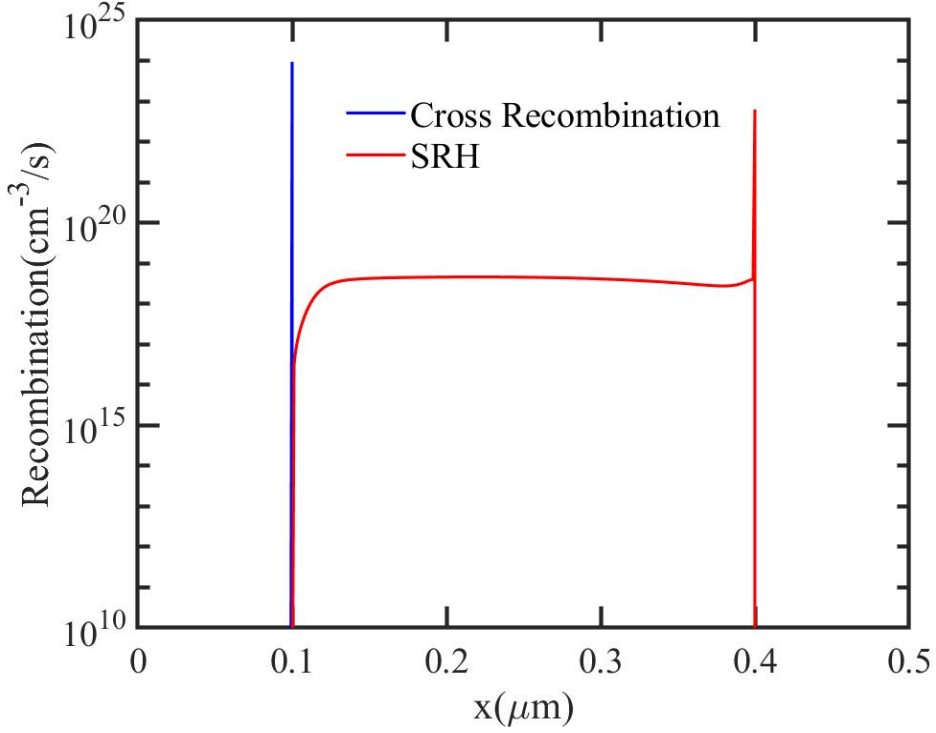

**FIG. S5.** Recombination profile of TPSC with and without cross recombination at $V_{bias} = V_{OC} - 85mV$. The band-band recombination due to eq. (6) is denoted as cross - recombination. It is evident that significant efficiency loss happens due to this recombination.

**Section VI: Derivation of dark current in TPSC**

The dark current in the TPSC is solely contributed by electrons as there isn't any source of holes, hence dark current is electron current itself. The continuity equation for electrons assuming no generation and no recombination is,

$$\frac{1}{q}\frac{\partial J_e}{\partial x} = 0 \qquad (S1)$$

Where, $q$ is the charge, $x$ is spatial coordinate and $J_e$ is given by,

$$J_e = q(n\mu_n\xi + D_n\frac{dn}{dx})$$

$$= qD_n\left(-\frac{n}{V_T}\frac{d\psi}{dx} + \frac{dn}{dx}\right) \qquad (S2)$$

Where $\psi$ is the potential, $V_T$ is the thermal voltage, $n$ is the electron density, $\xi$ is the electric field, $D_n$ is the diffusion coefficient and $\mu_n$ is the electron mobility. Note that the device with ions has lower $V_{OC}$ than the device without ions. However, we expect the trends to be same in both cases. For simplicity we assume a system without ions. Accordingly, one can assume linear electric field inside the active layer. Further, (eq. S1) suggests that the electron current density is spatially constant. Hence (eq. S2) can be rewritten as[3],

$$J_e = \frac{\int_0^W qD_n\left(-\frac{n}{V_T}\frac{d\psi}{dx} + \frac{dn}{dx}\right)e^{\frac{-\psi}{V_T}}dx}{\int_0^W e^{\frac{-\psi}{V_T}}dx}$$

$$= \frac{\int_0^W qD_n\frac{d}{dx}\ ne^{\frac{-\psi}{V_T}}dx}{\int_0^W e^{\frac{-\psi}{V_T}}dx} \qquad (S3)$$

The boundary condition for the device as per Fig. 1 of main manuscript is,

$$\psi(x) = \frac{(V_{BI} - V)x}{W}$$

$$n(0) = N_{TMO}e^{\frac{-V_{BI}}{V_T}}$$

$$n(W) = n(0)e^{\frac{V_{BI}}{V_T}}$$

Here, $N_{TMO}$ is the doping concentration of TMO layer which is nothing but $N_{dop}$ itself.

Substituting boundary conditions in (eq. S3) yields,

$$J_{dark} = J_e = -\frac{qN_{TMO}e^{\frac{-V_{BI}}{kT}}\mu_n(\frac{V_{BI} - V}{W})}{(e^{\frac{V - V_{BI}}{V_T}} - 1)}\ (e^{\frac{V}{kT}} - 1) \qquad (S4)$$

## Section VII: Estimation of open circuit voltage

The open circuit voltage of TPSC is the voltage (V) at which $J_{dark} = J_{SC}(= 25\ mA/cm^2)$. Hence eqn. S4 can be written as,

$$J_{SC} = -\frac{qN_{TMO}e^{\frac{-V_{BI}}{kT}}\mu_n\left(\frac{V_{BI}-V_{OC}}{W}\right)}{(e^{\frac{V_{OC}-V_{BI}}{V_T}}-1)}(e^{\frac{V_{OC}}{kT}}-1) \qquad (S5)$$

Table S2 shows the open circuit voltage values obtained from the above method.

| $V_{BI}(V)$ | Estimated $V_{OC}(V)$ Analytical method | Obtained $V_{OC}(V)$ From Numerical simulations |
|---|---|---|
| 1V | 0.653 | 0.688 |
| 1.1V | 0.752 | 0.788 |
| 1.2V | 0.853 | 0.885 |
| 1.3V | 0.953 | 0.968 |

**Table S2.** Estimated and actual $V_{OC}$ of TPSC with different $V_{BI}$

Further, for CPSC depending upon the value of surface recombination velocity ($S_v$) the dominant mechanism could be either SRH or surface recombination. Hence effective time constant ($\tau_{eff}$) is written as,

$$\tau_{eff} = [\frac{1}{\tau} + \frac{2S_v}{W}]^{-1} \qquad (S11)$$

And,

$$\Delta n = 2G\tau_{eff} \qquad (S12)$$

Accordingly, the open circuit voltage of CPSC is given by,

$$V_{OC} = 2kTln(\frac{2G\tau_{eff}}{n_i}) \qquad (S13)$$

Note that $\tau_{eff}$ is reduced from $10^{-6}s$ to $\sim 10^{-8}s$ due to surface recombination ($S_v = 10^3\ cm/s$) which yields a $V_{OC}$ of $\sim 0.98V$.

**Section VIII: Energy band diagram of TPSC in the presence of ions**

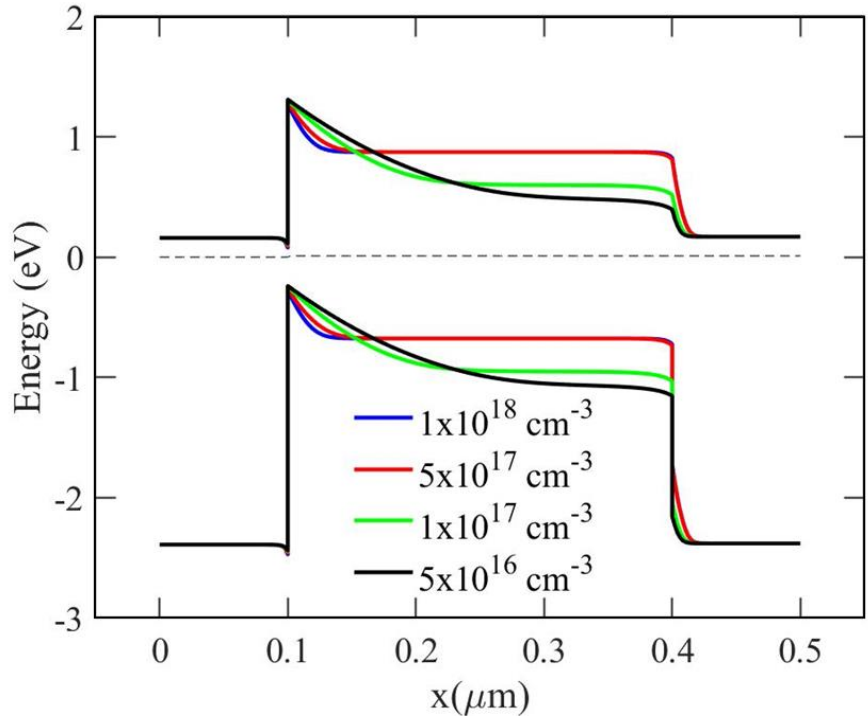

**FIG. S6.** Equilibrium energy band diagram of TPSC with different ion densities. The reduction of electric field inside the perovskite with an increase in ion density causes the efficiency droop as indicated in the Fig. 4 of main manuscript.

**Section IX: SRH profile near open circuit condition**

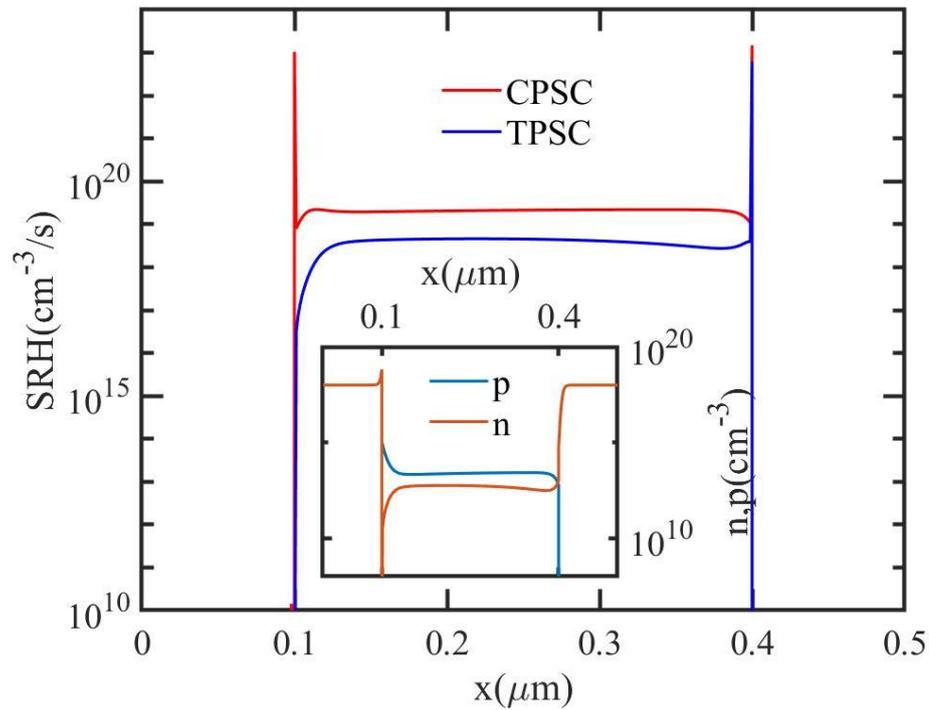

**FIG. S7.** SRH profiles for CPSC and TPSC near corresponding open circuit condition. The absence of SRH peak near TMO/Perovskite interface suggest than unlike CPSC interface degradation affects only one interface as discussed in the main manuscript. The inset shows the carrier density profile (electron(n), hole(p)) of TPSC near open circuit condition. Note that, the electron density inside perovskite near TMO interfcae is so low due to the upward band bending (see Fig. S4)in the perovskite which reduce the recombination.

**Section X: Effect of Ion density and Surface recombination velocity in TPSC**

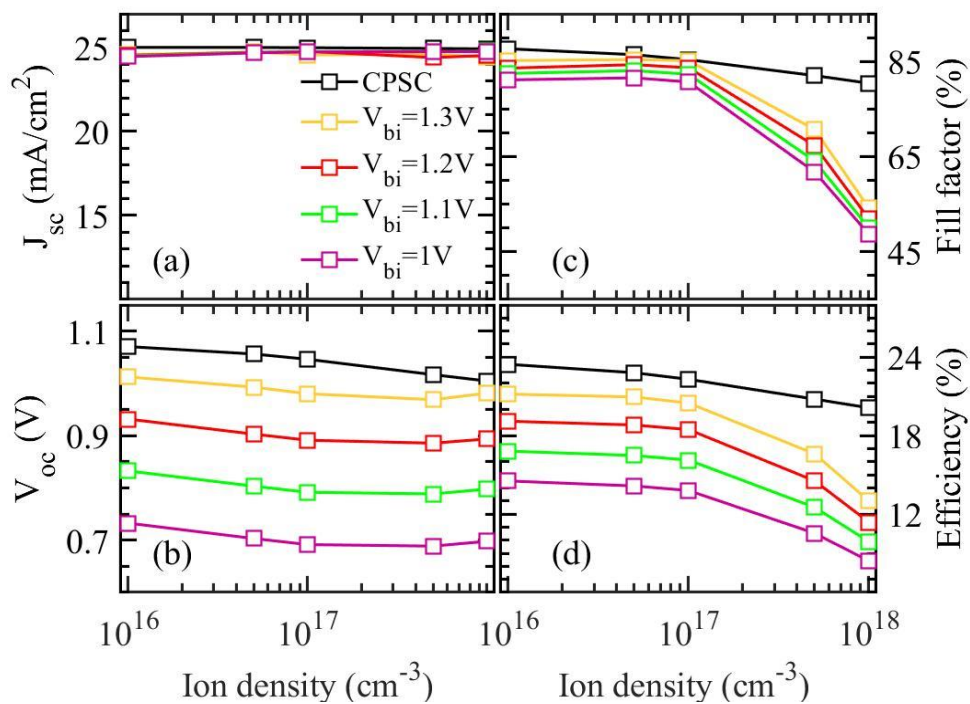

**FIG. S8.** Variation of Open circuit voltage(a), Short circuit current(b), fill factor(c) and efficiency(d) of CPSC ($V_{BI} = 1.2V$) and TMO-based solar cell with ion density.

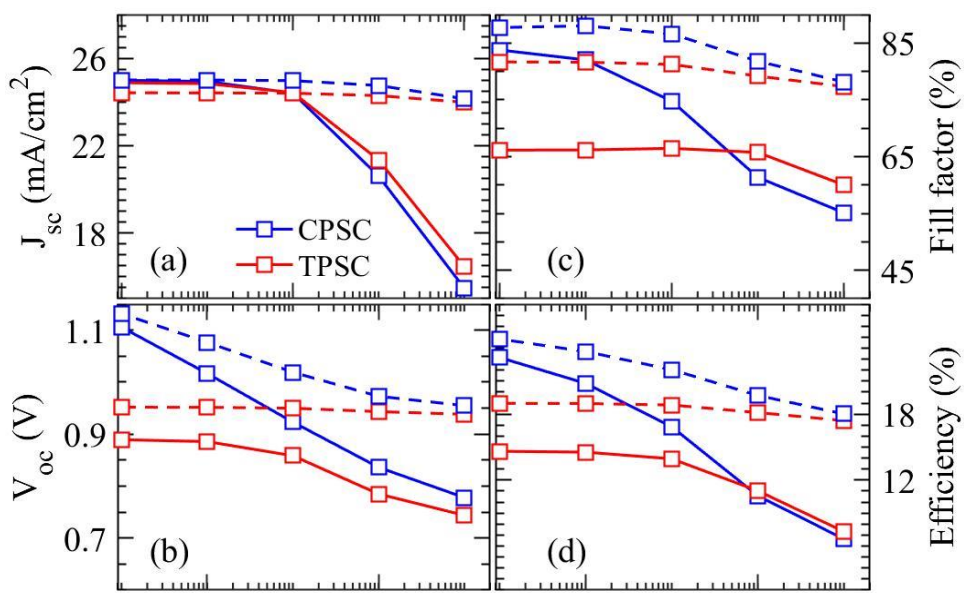

**FIG. S9.** Variation of Open circuit voltage(a), Short circuit current(b), fill factor(c) and efficiency(d) of CPSC and TPSC with surface recombination velocity ($S_v$). Dashed lines indicate devices in the absence of ions.

**Section XI: Effect of B parameter on JV characteristics of TPSC**

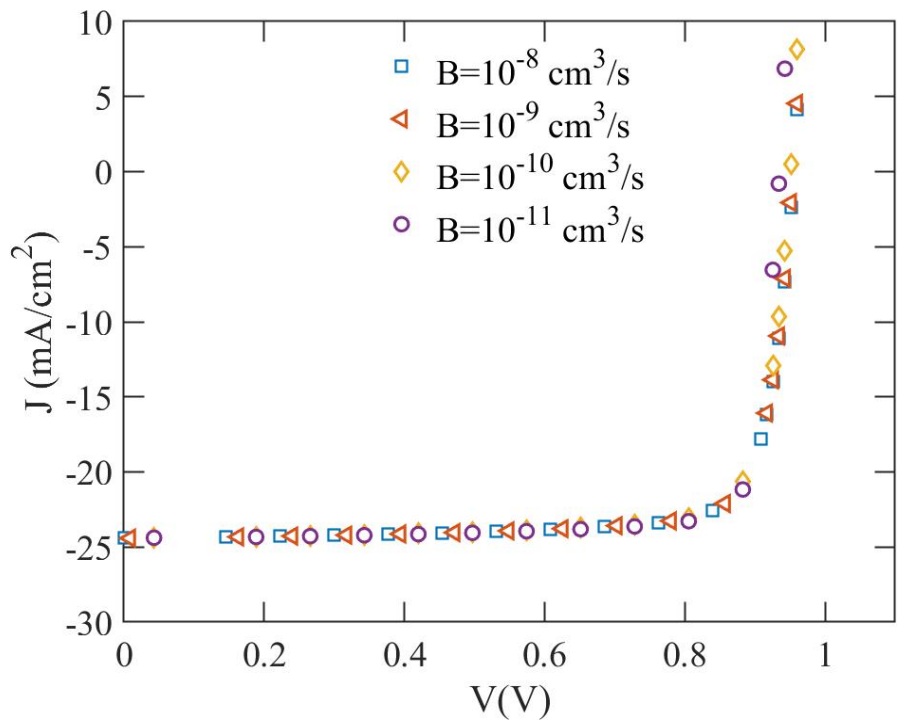

**FIG. S10.** JV Characteristics of TPSC ($V_{bi} = 1.2V$) with different B parameter. One could see B parmeter has mere effect on efficiency of TPSC.

**Section XII: JV characteristics of TPSC with efficiency of 21%**

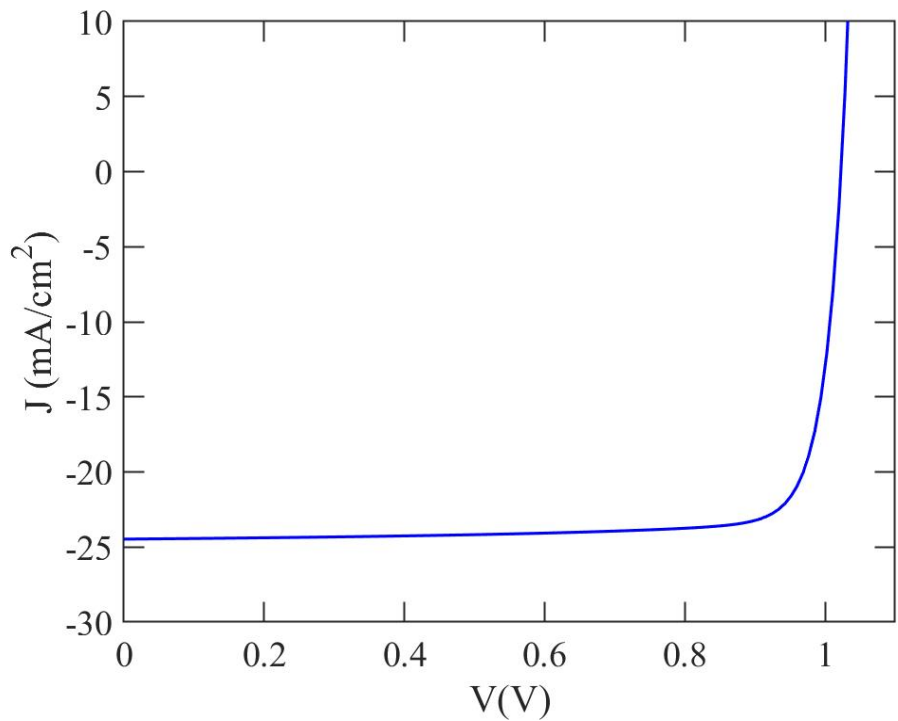

**FIG. S11.** JV Characteristics of TPSC ($V_{bi} = 1.3V$) with surface recombination velocity of $10 \; cm/s$. Device gives an efficiency of ~21%.